        \documentclass[10pt,journal,compsoc]{IEEEtran}
        \ifCLASSOPTIONcompsoc
          \usepackage[nocompress]{cite}
        \else
          \usepackage{cite}
        \fi
        \ifCLASSINFOpdf
        \else
        \fi
        \usepackage{graphicx}
        \usepackage{amsmath}
        \usepackage{multirow}
        \usepackage{booktabs}
        \usepackage{array}
        \usepackage{subfigure}
        \usepackage{amssymb}
        \usepackage{algorithm}
        \usepackage{verbatim}
        \usepackage{bm}     
        \usepackage{enumitem}       
        \usepackage{xcolor}
        \usepackage{algpseudocode}
        \usepackage{stfloats}
        \newcommand{\tabincell}[2]{\begin{tabular}{@{}#1@{}}#2\end{tabular}}
\begin{document}
        %
        \title{Personalized Prompt for Sequential Recommendation }

        \author{Yiqing~Wu,
                Ruobing~Xie,Yongchun~Zhu,
                Fuzhen~Zhuang$^*$,
                Xu~Zhang,
                Leyu~Lin~ and~
                Qing~He$^*$
        \IEEEcompsocitemizethanks{\IEEEcompsocthanksitem  Yiqing Wu, Yongchun Zhu, Qing He are with Institute of Computing Technology, Chinsese Academic of Science, Beijing 100190, China, and University of Chinese Academy of Sciences, Beijing 100049, China.
        E-mail: \{wuyiqing20s, zhuyongchun18s, heqing\}@ict.ac.cn
        \IEEEcompsocthanksitem  Ruobing Xie, Xu Zhang and Leyu Lin are with WeChat, Tencent, China.
        E-mail:  \{ruobingxie,xuonezhang,goshawklin\}@tencent.com
        \IEEEcompsocthanksitem Fuzhen Zhuang is with Institute of Artificial Intelligence, Beihang University, Beijing 100191, China, and SKLSDE, School of Computer Science, Beihang University, Beijing 100191, China.
        E-mail: zhuangfuzhen@buaa.edu.cn}
        
        \thanks{ ${*}$ indicates corresponding author.}}
        
        %
        %

        \markboth{TKDE}%
        {Shell \MakeLowercase{\textit{et al.}}: Bare Demo of IEEEtran.cls for Computer Society Journals}
        %



        \IEEEtitleabstractindextext{%
        \begin{abstract}
        Pre-training models have shown their power in sequential recommendation. Recently, prompt has been widely explored and verified for tuning after pre-training in NLP, which  helps to more effectively and parameter-efficiently extract useful knowledge from pre-training models for downstream tasks, especially in cold-start scenarios. However, it is challenging to bring prompt-tuning from NLP to recommendation, since the tokens of recommendation (i.e., items) are million-level and do not have concrete explainable semantics, and the sequence modeling in recommendation should be personalized. In this work, we first introduce prompt to  recommendation models and propose a novel Personalized prompt-based recommendation (PPR) framework for cold-start recommendation. Specifically, we build personalized soft prompt via a prompt generator based on user profiles, and enable a sufficient training on prompts via a new prompt-oriented contrastive learning.
        PPR is effective, parameter-efficient, and universal in various tasks. In both few-shot and zero-shot recommendation tasks, PPR models achieve significant improvements over baselines in three large-scale datasets.
        We also verify PPR's universality in adopting different recommendation models as the backbone. Finally, we explore and confirm the capability of PPR on other tasks such as cross-domain recommendation and user profile prediction, shedding lights on the promising future directions of better using large-scale pre-trained recommendation models.
        \end{abstract}
        
        }

        \maketitle

        \IEEEdisplaynontitleabstractindextext

        %
        \IEEEpeerreviewmaketitle

        \IEEEraisesectionheading{\section{Introduction}
        \label{sec.introduction}}
        

        %
        %
        %
        %
        
        \IEEEPARstart{P}{ersonalized} recommendation aims to provide appropriate items for users according to their preferences, where user historical behavior sequence is an informative source for user understanding. \emph{Sequential recommendation}, which takes users' historical behavior sequences as inputs and outputs the next predicted items, is widely studied and deployed in practice \cite{kang2018self,sun2019bert4rec,xie2021adversarial,xie2022contrastive}.
        With the thriving of \emph{pre-training} in NLP \cite{devlin2019bert}, there are lots of efforts that bring pre-training into sequential recommendation \cite{zeng2021knowledge}. These pre-trained recommendation models usually consider user behavior sequences as token sequences in NLP, using pre-training techniques to improve the sequence modeling ability of user behaviors, which can alleviate the sparsity issues in real-world recommendation systems \cite{xiao2021uprec}.
        
        
        Recently, with the overwhelming trend of pre-training, how to effectively and efficiently extract useful information from large-scale pre-trained models has become a promising direction. \textbf{Prompt-tuning} \cite{brown2020language,lester2021power,han2022ptr} is a representative and powerful method that has remarkable superiority over the classical fine-tuning paradigm, especially in zero-shot and few-shot scenarios. It usually inserts hard text templates \cite{brown2020language,gao2021making} or soft continuous embeddings \cite{li2021prefix,qin2021learning} as \textbf{prompts}, and transforms the downstream tasks into similar well-trained pre-training tasks. The advantages of prompt-tuning locate in two aspects:
        (1) it bridges the gap between pre-training and downstream objectives, which could better utilize the knowledge in pre-training models. This advantage will be multiplied in cold-start scenarios.
        (2) Prompt-tuning only needs to tune a small set of parameters for the prompts and labels, which is more parameter-efficient.
        Looking back to the sequential recommendation task, the cold-start user issues (including zero-shot and few-shot scenarios) are crucial challenges due to the sparsity of interactions in real-world systems. In this work, with pre-training techniques widely adopted to address cold-start issues, we attempt to bring in prompt-tuning to better extract informative knowledge from the huge pre-trained recommendation models.
        
        
        \begin{figure}[!hbtp]
        \centering
        \includegraphics[width=0.99\columnwidth]{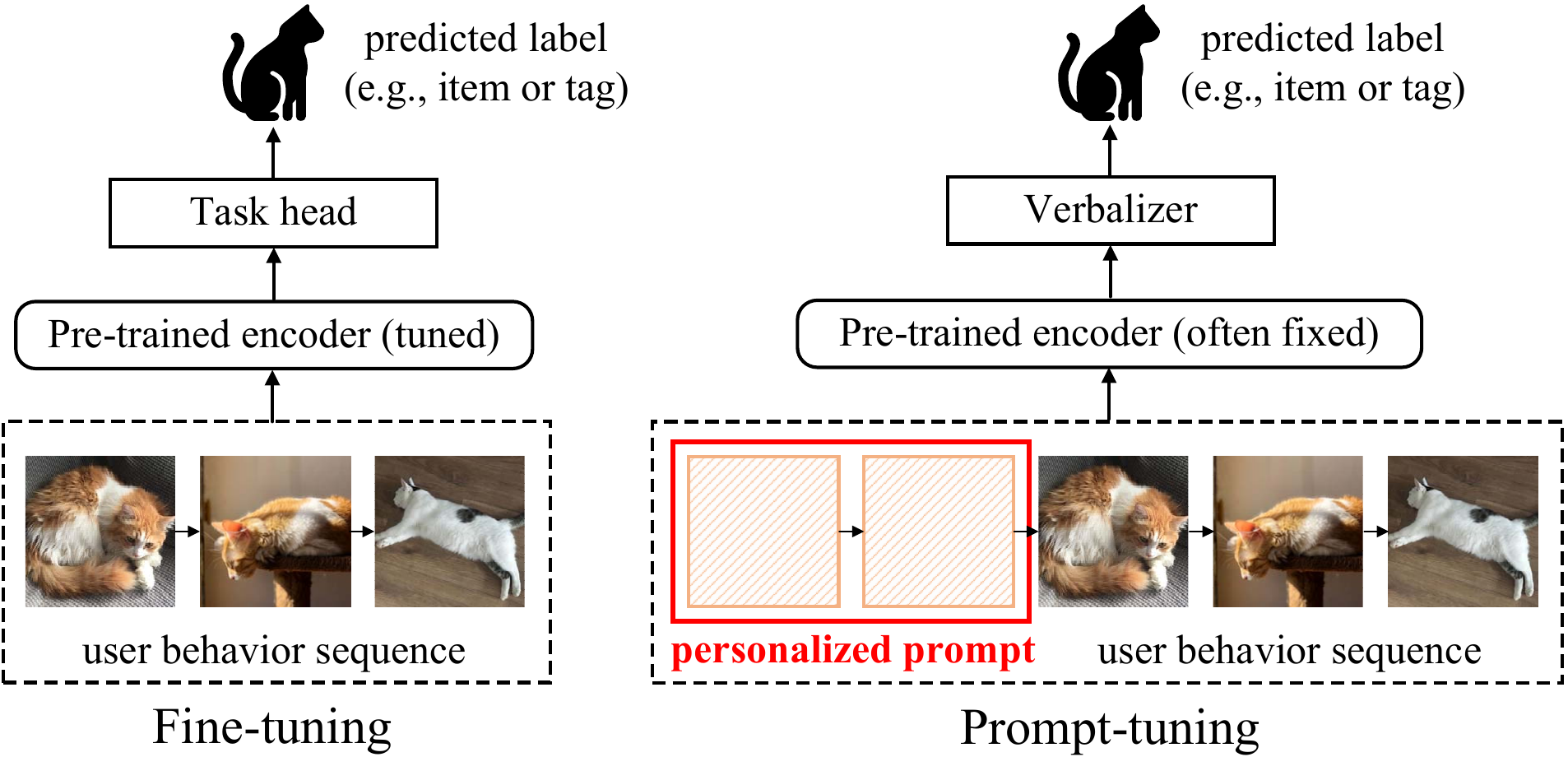}
        \caption{An example of prompt-tuning in recommendation.}
        \label{fig:example}
        \end{figure}
        
        However, adopting prompts in recommendation is non-trivial due to the following challenges:
        (1) \emph{\textbf{How to transfer the prompt-tuning paradigm from NLP into recommendation}}? Differing from words in NLP, items in recommendation do not have concrete semantic meanings, and thus cannot be directly used to build hard explainable prompts to prompt pre-trained models. Moreover, it is also challenging to design an appropriate framework to make full use of pre-training knowledge for various recommendation tasks.
        (2) \emph{\textbf{How to construct effective personalized prompts for recommendation}}? Compared with NLP, recommendation further emphasizes personalization. Hence, the proposed prompts should better be personalized so as to more pertinently extract user-related knowledge from the huge pre-trained recommendation models to downstream tasks.
        
        
        To address these challenges, we propose a novel \textbf{Personalized prompt-based recommendation (PPR)} framework, which first adopts prompt-tuning in recommendation models. Specifically, PPR adopts the \emph{personalized soft prefix prompt} learned via our prompt generator based on user information such as user profiles, which elegantly prompts the pre-trained models to provide user-related knowledge for various downstream recommendation tasks within a universal tuning framework. We also propose a \emph{prompt-oriented contrastive learning} via both prompt-based and behavior-based data augmentations to further enhance the training of prompts against data sparsity.
        Compared with fine-tuning, PPR has the following advantages:
        (1) PPR enables the prompt-tuning of recommendation models, making the usage of pre-training in downstream tasks more effective, personal and parameter-efficient.
        (2) PPR designs a set of prompt-oriented contrastive learning losses on prompt-enhanced behavior sequences, which enables a more sufficient training for prompts.
        (3) PPR is effective, parameter-efficient, universal, and easy-to-deploy, which can be conveniently adopted on various recommendation models and downstream tasks such as cross-domain recommendation and user profile prediction.
        
        
        In experiments, we conduct extensive evaluations to verify the effectiveness and universality of our personalized prompt-based recommendation. We conduct evaluations on three large-scale datasets in few-shot and zero-shot scenarios, which confirms that PPR can achieve significant improvements in both scenarios. Ablation studies and sparsity analyses are also conducted. Moreover, we also verify PPR's universality on multiple pre-training recommendation models (e.g., SASRec \cite{kang2018self} and CL4SRec \cite{xie2022contrastive}) and other challenging downstream tasks (e.g., cross-domain recommendation and user profile prediction), shedding light on the promising and wide applications of personalized prompts in practice. The contributions of this work are as follows:
        \begin{itemize}
         \item We propose a novel personalized prompt-based recommendation framework for various recommendation tasks. To the best of our knowledge, we are the first to adopt prompt-tuning for better extracting knowledge from pre-trained recommendation models.
         \item We design the novel prompt-oriented contrastive learning considering both prompt- and behavior- based augmentations, enabling more sufficient training.
          \item We have verified the effectiveness and universality of PPR on multiple backbone models and tasks including cold-start recommendation, cross-domain recommendation, and user profile prediction. It implies the promising possibility of adopting large-scale pre-trained recommendation models with prompt-tuning as a new recommendation paradigm in the future.
        \end{itemize}

        \section{Related Works}
        \label{sec.related_work}
        
        \noindent
        \textbf{Sequential Recommendation.}
        Sequential recommendation models mainly leverage users' chronological behavior sequences to learn users' preferences.\cite{hidasi2016parallel,vaswani2017attention,devlin2019bert,kang2018self,sun2019bert4rec,zheng2021cold,lian2020geography,liu2021augmenting,han2023guesr}
        Recently, various deep neural networks have been employed for sequence-based recommendation. GRU4Rec \cite{hidasi2016session} proposes to use Gated Recurrent Units in the session-based recommendation. Inspired by the success of Transformer  and BERT \cite{vaswani2017attention,devlin2019bert}, SASRec \cite{kang2018self} and Bert4Rec \cite{sun2019bert4rec} adopt self-attention mechanisms to model user behavior sequences. The cold-start problem on sequential recommendation also attracts the wide attention of researchers. \cite{zheng2021cold} utilizes a meta-learning mechanism to alleviate  cold-start item problems in sequential recommendation. \cite{liu2021augmenting} augments short behavior sequence by reversely predicted items. Pre-training \& fine-tuning is also a widely-used method for cold-start recommendation \cite{zeng2021knowledge,chen2021user}.
        In this work, we focus on improving the pre-training \& fine-tuning manner.
        
        
        \noindent
        \textbf{Pre-training in Recommendation.} Recently, pre-training models have achieved great success in NLP \cite{devlin2019bert,liu2019roberta} and CV \cite{he2021masked,radford2021learning}. It aims to learn prior knowledge from general large-scale datasets to help the specific downstream tasks. After pre-training, models are further fine-tuned on downstream supervised signals to fit the specific task. This \emph{pre-training and fine-tuning} paradigm is widely applied to various tasks \cite{devlin2019bert}.
        With the thriving of a pre-training, many pre-training models have been proposed in recommendation\cite{shin2021one4all,yao2021self,wu2021self,huang2023recommender}. BERT4Rec \cite{sun2019bert4rec} adopts masked item prediction in sequential recommendation. $S^3$Rec \cite{zhou2020s3} pre-trains the sequential encoder considering the correlations among item’s attributes, item, subsequence, and sequence. CL4SRec \cite{xie2022contrastive} applies contrastive learning (CL) via item crop, mask, and reorder on sequence modeling. PeterRec\cite{yuan2020parameter} transfers pre-trained model to solve cross-domain recommendation via Adapters. UPRec \cite{xiao2021uprec} further highlights user profiles and social relations in pre-training.
        
        Recently, some works attempt to introduce pre-trained language model (PLM) into recommendation. UniSRec \cite{hou2022towards} utilizes PLM to learn universal item representations across different domains via items' textual information. PEPLER \cite{li2022personalized} adopts users'/items' textual attributes as prompts of PLM to generate textual explanations for recommendations. P5 \cite{geng2022recommendation} follows PEPLER and designs various prompts to jointly learn from PLM for multiple downstream recommendation tasks. M6-Rec \cite{cui2022m6} also converts user behaviors and user/item attributes into textual data and models them via PLM.
        Different from these models that rely on \textbf{pre-trained language models} and mainly convert sequential behaviors into texts, our PPR concentrates on the intrinsic \textbf{pre-trained recommendation models} and their learned \emph{behavioral} knowledge. Moreover, PPR does not depend on additional textual information and PLMs, and thus is more universal and flexible. 
        Inspired by the successes of pre-training, we propose PPR, which aims to (a) narrow the gap between pre-training and downstream recommendation tasks, and (b) better extract useful personalized knowledge from pre-trained models by replacing fine-tuning with personalized prompt-tuning. We have also deployed PPR on different tasks and pre-training models to verify its effectiveness.

        
        
        
        \noindent
        \textbf{Prompt Tuning.}
        Prompt-tuning is first proposed in NLP, and is widely explored and dominating, especially in few-shot scenarios \cite{li2021prefix,radford2019language,liu2021gpt,liu2022p}.
        \cite{schick2020exploiting} and \cite{brown2020language} adopt hard prompts that consist of discrete real words. Considering that manually designing the hard prompt is both time-consuming and trivial, other works \cite{gao2021making,jiang2020can} focus on automatically searching for hard prompts.
        In contrast, soft prompts are composed of several continuous learnable embeddings randomly initialized.
        Prefix-tuning \cite{li2021prefix} optimizes continuous prompts for generation tasks in NLP.
        \cite{wu2022selective} adopts prefix prompts for selective fairness.
        PPT \cite{gu2021ppt} further conducts pre-training on the prompts to better initialize soft prompts. 
        In PPR, we first introduce prompts to better utilize pre-trained recommendation models. Different from prompts in NLP, we design personalized prompts.
        
        
        \section{Methodology}
        \label{sec.method}
        
        \subsection{Preliminaries}
        \label{sec.preliminaries}
        
        \begin{figure*}[!hbtp]
        \centering
        \includegraphics[width=0.98\textwidth]{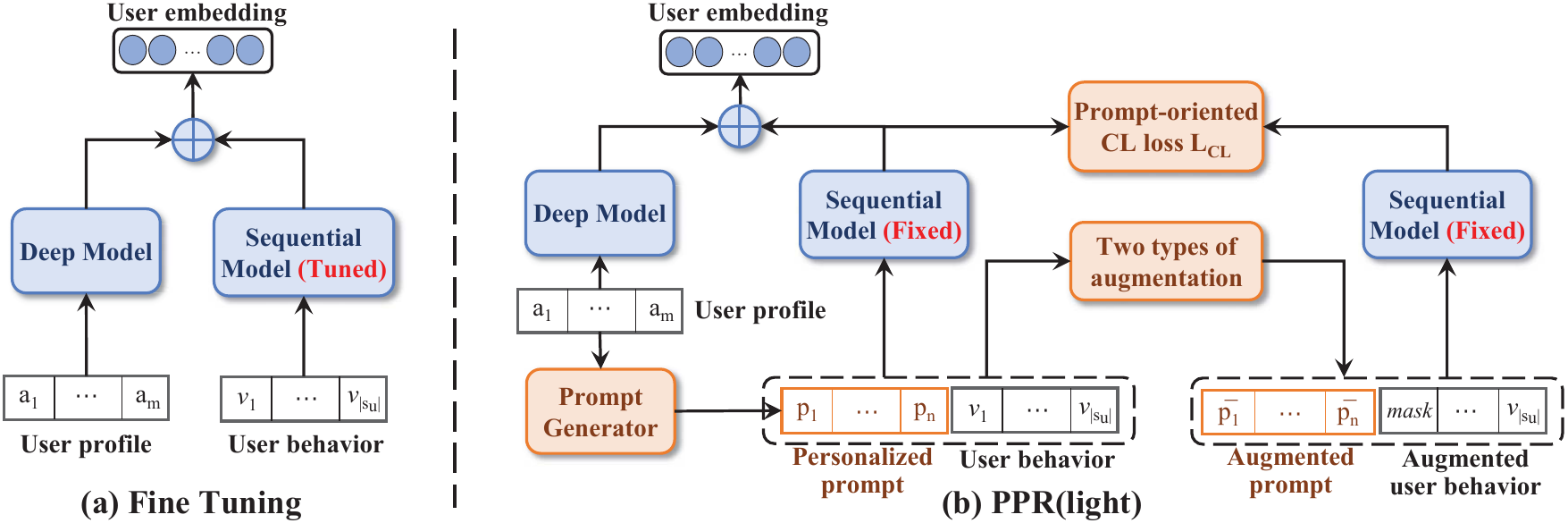}
        \caption{Overall architecture of  (a) the conventional fine-tuning, and (b) our Personalized prompt-based recommendation.}
        \label{fig:overall}
        \end{figure*}
        
        \subsubsection{Background of Prompt in NLP}
        
        In NLP pre-training, prompts are often a piece of natural language inserted into the input text sequence. For example, in sentiment analysis, a prompt ``it is [mask]'' is inserted after the review ``a real joy'' as a natural sentence: ``a real joy, it is [mask]''. In prompt-tuning, the predicted tokens of ``[mask]'' (e.g., great) will be mapped to the sentiment labels (e.g., positive) via a task-specific verbalizer. In this case, the original task can be formulated as a masked language model task, which has been fully optimized in pre-training and thus has better performance \cite{radford2019language,brown2020language,shin2020autoprompt}.
        The advantages of prompt-tuning are as follows: (1) it can better extract useful knowledge in pre-trained models via prompts based on similar modeling and objectives fully learned in pre-training, especially for few-shot scenarios compared to fine-tuning \cite{gu2021ppt}, and (2) it is more parameter-efficient compared to fine-tuning, for the tuned parameters in prompt-tuning are far fewer than those of pre-trained models.
        
        \subsubsection{Goals of PPR}
        
        In this work, we propose Personalized prompt-based recommendation to provide a more effective and parameter-efficient tuning for (cold-start) downstream tasks based on pre-trained sequential recommendation models.
        The \emph{pre-training and fine-tuning} is a classical and widely-used training paradigm, which first trains a pre-trained model on a large-scale general dataset, and then fine-tunes the whole pre-trained model by the supervised information of downstream tasks.
        Recently, the \textbf{pre-training and prompt-tuning} paradigm has been widely verified in NLP as in Sec. \ref{sec.related_work}, where the prompts and verbalizer are mainly updated for efficiency. PPR attempts to better extract useful personalized information from pre-trained models via prompt-tuning rather than fine-tuning, and thus fine-tuning is the most essential baseline.
        Precisely, the main task that our PPR focuses on is cold-start recommendation (since prompt-tuning functions well in few-shot learning), which consists of both \emph{few-shot recommendation} and \emph{zero-shot recommendation}. Moreover, we also deploy PPR for other downstream tasks such as \emph{cross-domain recommendation} and \emph{user profile prediction} in Sec. \ref{sec.explorations} to analyze PPR's universality.
        
        \subsubsection{Notions of PPR}
        
        The key notions of PPR are defined as follows. We denote user and item as $u\in U$ and $v\in V$. $U$ and $V$ are the overall user and item sets. Each user $u$ has a historical behavior sequence $s_u=\{v_1^{u},v_2^{u},...,v_{|s_u|}^{u}\}$ (of length $|s_u|$) ordered by time. Each user has $m$ attributes $A_u=\{a_1^u,a_2^u,...,a_m^u\}$ (i.e., user profiles such as age and gender).
        In cold-start recommendation, we split users into a warm user set $U^W$ (for pre-training) and a cold user set $U^C$ (for tuning and evaluation) by their behavior sequence length.
        Then, we first train a sequential model $f_{seq}(\{v_1^u,v_2^u,...,v_{|s_u|}^u\}|\Theta)$ to learn user representations by warm users' historical behaviors called warmed model \footnote{In this study, the warmed model is similar with the pre-trained model. However, considering that mainstream pre-trained models are trained across multiple domains and applied in various fields, in order to avoid ambiguity with mainstream pre-trained models, we refer to the trained model as warmed model.}, where $\Theta$ is the warmed model parameters. After that, we fine-tune the pre-trained model on cold users' historical behaviors and get the fine-tuned model $f_{seq}(\{v_1^u,v_2^u,...,v_{|s_u|}^u\}|\hat{\Theta})$, where $\hat{\Theta}$ is the fine-tuned model's parameters. Different from fine-tuning, our PPR inserts personalized prompts (i.e., prefix tokens) before behavior sequences, noted as $f_{seq}(\{p_1^u,p_2^u,...,p_n^u,v_1^u,v_2^u,...,v_{|s_u|}^u\}|\Theta,\vartheta)$, where $\{p_1^u,p_2^u,..,p_n^u\}$ is the personalized
        prompt built from user profiles, and $\vartheta$ is the prompt generator's parameters.
        

        
        
        \subsection{Overall Framework}
        \label{sec.overall}
        
        The overall framework of PPR is illustrated in Fig. \ref{fig:overall}. For each user, we first build the personalized prompts according to user profiles via the prompt generator, and insert them into the beginning of the user behavior sequence. The prompt-enhanced sequence is then fed into the warmed sequential model to generate user's behavioral representations. Moreover, besides the prompt generator, the user profiles are further fed into another Deep model to generate user's attribute-level representations. Finally, the user's behavioral and attribute-level preferences are combined to get the final user representation. To enable a more sufficient training of personalized prompts, we further design a prompt-oriented contrastive learning loss, adopting two augmentations on both the prompt generator and sequence modeling parts.
        We design two PPR versions for different practical demands. In PPR(light), the warmed model is fixed during tuning as classical prompt-tuning in NLP, which is more parameter-efficient. While in PPR(full), all prompts and warmed parameters are tuned specially for the characteristics of recommendation.
        
        \subsection{Warm-training Stage of PPR}
        \label{sec.pre-training}
        
        We first introduce the warm-training stage of PPR. There are various self-attention based sequential models \cite{kang2018self,sun2019bert4rec,xie2022contrastive} verified to be effective as base models. Following \cite{xie2022contrastive}, we also use the classical SASRec \cite{kang2018self} as our base model. Specifically, SASRec stacks Transformer(·) blocks to encode the historical behavior sequence. For the input behavior sequence $s_u$, we define its $l$-layer's behavior matrix as $\bm{H}_u^l=\{\bm{h}_{u,1}^l,\bm{h}_{u,2}^l, \cdots, \bm{h}^{l}_{u,|s_u|}\}$, where $\bm{h}_{u,i}^l$ is the $i$-th behavior's representation of $u$ at the $l$-th layer. The $(l+1)$-layer's behavior matrix $\bm{H}_u^{l+1}$ is then learned as follows:
        \begin{equation}
        \begin{aligned}
        \bm{H}_u^{l+1}=\mathrm{Transformer}^l(\bm{H}_{u}^{l}), 
        \\
        \bm{u}_o=f_{seq}(s_u|\Theta)=\bm{h}_{u,|s_{u}|}^{L}.
        \end{aligned}
        \label{eq.transformer}
        \end{equation}
        $\bm{u}_o$ is the final user representation of $u$ learned in pre-training to predict the user's next items, which is generated as the last behavior's representation at the $L$-th layer (i.e., $\bm{h}_{u,|s_{u}|}^{L}$). Here, $L$ is the number of Transformer layers.
        
        Following classical ranking models \cite{rendle2009bpr,kang2018self,xie2022contrastive}, the pre-training model is optimized under the objective $L_o$ as:
        \begin{equation}
        \begin{split}
        L_o = -\sum_{(u,v_i) \in S^w_+} \sum_{(u,v_j) \in S^w_-} \log \sigma (\bm{u}_o^\top\bm{v}_i - \bm{u}_o^\top\bm{v}_j), \quad u\in U^w.
        \end{split}
        \label{eq.simple_loss}
        \end{equation}
        $(u,v_i) \in S^w_+$ indicates the positive set where $u$ has clicked $v_i$,  and $(u,v_j) \in S^w_-$ indicates the negative set where $v_j$ is randomly sampled negative items. $\sigma (\cdot)$ is the sigmoid function. We optimize the parameters $\Theta$ of the based model via $L_o$ on the pre-training dataset of $u \in U^w$.
        Differing from classical prompt-tuning in NLP, which usually adopts the masked language model (MLM) pre-training task \cite{devlin2019bert}, PPR mainly concentrates on the next-item prediction task in pre-training. It is because the next-item prediction task perfectly fits our downstream tasks (cold-start recommendation and user profile prediction) with PPR, which can stimulate the maximum potential of prompt.
        Note that our PPR can be flexibly adopted on different models as the backbone. We verify PPR's universality on CL4SRec \cite{xie2022contrastive} in Sec. \ref{sec.universality}.
        
        \subsection{Personalized Prompt-tuning}
        
        After warm-training, PPR conducts a personalized prompt-tuning instead of fine-tuning to better extract useful information from warmed models to downstream tasks.
        
        \subsubsection{Personalized Prompt Generator}
        
        The key of our PPR is generating an effective prompt that helps to narrow the gap between pre-trained models and downstream tasks. However, it is challenging to find appropriate prompts in recommendation, since (1) it is difficult to build hard prompts (i.e., some real tokens) in PPR. Unlike words in NLP, the tokens (i.e., items) in recommendation do not have explicit meaningful semantics. (2) Moreover, unlike NLP, recommendation should be personalized. thus the prompts should also be customized for different users. In a sense, each user's recommendation can be viewed as a task, while there are millions of users in a real-world system. It is impossible to design prompts manually for each user.
        
        In PPR, to automatically and effectively build personalized prompts for all users, we rely on essential and informative user profiles. \emph{User profiles} can be produced by all types of user information, such as user static attributes (e.g., age, gender, location), user cumulative interests, and user behaviors in other domains. In the task of cold-start recommendation, we mainly consider the user static attributes $A_u$ to build prompts. Specifically, we concatenate $m$ user profile embeddings as $\bm{x}_u=[\bm{a}_1^u||\bm{a}_2^u|| \cdots ||\bm{a}_m^u]$, where $\bm{a}_i^u$ is the $i$-th user profile embedding. We conduct a Multi-layer perceptron (MLP) to learn the personalized prompt representation $\bm{P}^u=\{\bm{p}_1^u, \cdots, \bm{p}_n^u\}$ containing $n$ tokens as:
        \begin{equation}
        \begin{split}
            \bm{P}^u  = \mathrm{PPG} (\bm{x}_u|\vartheta)=\bm{W}_2\sigma(\bm{W}_1\bm{x}_u+\bm{b}_1)+\bm{b}_2,
        \end{split}
        \label{eq.prompt}
        \end{equation}
        where $\bm{W}_1\in \mathbb{R}^{d_1 \times d'}$ ,$\bm{W}_2\in \mathbb{R}^{d'\times d_2\times n}$, $\bm{b}_1\in \mathbb{R}^{d'}$ and $\bm{b}_2 \in \mathbb{R}^{d_2 \times n}$ are trainable parameters in $\vartheta$. $d_1$, $d'$, and $d_2$ are the embedding sizes of concatenated user profiles $\bm{x}_u$, hidden layer, and output prompts respectively. $n$ is the number of prompt tokens. We choose the current $\mathrm{PPG}(\cdot)$ function since it is simple and effective. It is also convenient to adopt other neural networks to build the prompts $\bm{P}^u$ from $\bm{x}_u$.
        
        Inspired by the success of prefix-tuning \cite{li2021prefix},  we adopt the prompt tokens as a prefix and connect them with user behavior sequence as:  $\hat{s}_u=\{p_1^u,p_2^u,...,p_n^u,v_1^u,v_2^u,...,v_{|s_u|}^u\}$. This \emph{prompt-enhanced sequence} contains both task-specific and user-specific information. For example, if $A_u=\{$20-year-old, female$\}$, the prompt-enhanced sequence $\hat{s}_u$ can be translated as ``In cold-start recommendation, a user is a 20-year-old female, she likes $v_1$, $v_2$, $\cdots$ '', which provides a more personalized context of user behavior sequences that better prompts the sequential model to give appropriate recommendations. 
        Different from the side Deep model of Fine-tuning in Fig. \ref{fig:overall}(a) that captures user profile information separately from user behaviors, our PPR naturally fuses heterogeneous user profiles into user historical behaviors, and models them via a joint sequential modeling inherited from the warmed recommendation model.
        In this case, the user-related knowledge hidden in warmed recommendation models can be fully activated for downstream tasks, which is essential in cold-start scenarios.
        
        \subsubsection{Prompt-tuning of PPR}
        \label{sec.tuning_PPR}
        
        After generating the personalized prompts, we need to construct the final user representation for recommendation. Precisely, we first input the prompt-enhanced sequence $\hat{s}_u$ to the warmed sequential model in Eq. (\ref{eq.transformer}) to get the user behavioral preference $\bm{u}_s$. Similar to fine-tuning in Fig. \ref{fig:overall}, we also use a Deep model (MLP) to learn the user attribute preference $\bm{u}_a$ from user profiles. Finally, both $\bm{u}_s$ and $\bm{u}_a$ are combined to get the final user representation $\bm{u}_p$. We have:
        \begin{equation}
        \begin{split}
        \bm{u}_p = \bm{u}_a + \bm{u}_s, \quad
        \bm{u}_a = \mathrm{MLP}_a (\bm{x}_u|\phi), \quad
        \bm{u}_s = f_{seq}(\hat{s}_u|\Theta,\vartheta).
        \end{split}
        \label{eq.prompt_user}
        \end{equation}
        $\Theta$, $\vartheta$, and $\phi$ are parameters of the warmed sequential model, the prompt generator, and the Deep model for $\bm{u}_a$, respectively. Finally, we follow the pre-training objective in Eq. (\ref{eq.simple_loss}) to build the optimization objective $L_p$ of our prompt-tuning for col-start recommendation as follows:
        \begin{equation}
        \begin{split}
        L_p = -\sum_{(u,v_i) \in S^c_+} \sum_{(u,v_j) \in S^c_-} \log \sigma (\bm{u}_p^\top\bm{v}_i - \bm{u}_p^\top\bm{v}_j), \quad u\in U^c.
        \end{split}
        \label{eq.prompt_loss}
        \end{equation}
        $S^c_+$ and $S^c_-$ are the same positive and negative sample sets in fine-tuning. We adopt similar training objectives for warm-training and prompt-tuning, aiming to better use pre-trained knowledge for downstream tasks via parameter-efficient prompt-tuning.
        
        To tune these parameters, we propose two prompt tuning strategies, PPR(light) and PPR(full), to balance effectiveness and efficiency for practical demands.
        
        \textbf{PPR(light).}
        In PPR(light), we merely update the newly-introduced parameters, i.e., $\vartheta$ of the prompt generator in Eq. (\ref{eq.prompt}) and $\phi$ of the user profile learner in Eq. (\ref{eq.prompt_user}), with other parameters fixed. It is similar to the typical prompt-tuning manner in NLP \cite{li2021prefix}. This is a straightforward and efficient prompt-tuning manner, which completely relies on the warmed recommendation model in sequence modeling and item representation learning. The tuned parameters of PPR(light) are greatly reduced compared to fine-tuning (note that even the item embeddings to be predicted are also fixed for efficiency). The tuned personalized prompts work as an inducer, which smartly extracts useful knowledge related to the current user from a large-scale pre-trained model.
        
        \textbf{PPR(full).}
        However, due to the huge gaps between NLP and recommendation tasks, the widely-verified original ''light'' prompt-tuning does not always perform well enough in downstream tasks. It is because that (a) plenty of NLP downstream tasks (e.g., sentiment analysis, text classification) only have limited numbers of predicted labels, and (b) high-frequent words are often fewer than several thousand. On the contrary, items (also labels to predict) in recommendation are often million-level in practice, while they are fixed in PPR(light).
        Hence, we propose another PPR(full) manner for more comprehensive tuning. Specially for a recommendation, which further tunes the warmed model's parameters $\Theta$ (including parameters of item embeddings and sequential model) besides $\vartheta$ and $\phi$. In a sense, PPR(full) can be regarded as  prompt-enhanced tuning by combining fine-tuning and PPR(light).
        
        
        \subsection{Prompt-oriented Contrastive Learning}
        \label{sec.data_augmentation}
        
        The main challenge of cold-start recommendation is the lack of sufficient tuning instances and the noise existing in cold user behaviors, and similar issues also exist in prompt-tuning. 
        Recently, contrastive learning (CL) has shown its power in recommendation \cite{xie2022contrastive,zhou2020s3}. These CL-based models usually conduct self-supervised learning (SSL) as supplements to supervised information via certain data augmentations, which could obtain more effective and robust user representations and alleviate the data sparsity issues. Inspired by those methods, we also adopt CL as auxiliary losses via two types of data augmentations based on our prompt-enhanced sequences.
        
        \subsubsection{Prompt-based Augmentation}
        
        In real-world systems, users' attributes are usually noisy or even missing, while they are the main source of our personalized prompts, whose qualities are essential, especially in zero-shot scenarios. Therefore, we design a prompt-based data augmentation to improve the effectiveness and robustness of the prompt generator. Specifically, we conduct random element-level masking on the feature elements of user profile embeddings $\bm{x}_u$ to obtain $\bm{\bar{x}}_u$ with a certain mask ratio $\gamma_1$. Formally, the augmented prompt-enhanced behavior sequence is noted as $\bar{s}_u^1=\{\bar{p}_1^u,\bar{p}_2^u, ...,\bar{p}_n^u,v_1^u,v_2^u, ...,v_{|s_u|}^u\}$, where $\bm{\bar{P}}^u  = \mathrm{PPG} (\bm{\bar{x}}_u|\vartheta)$ is learned from masked attributes. This augmentation can also avoid the possible overfitting of our prompt generator on some abnormal profiles.
        
        \subsubsection{Behavior-based Augmentation}
        
        Besides the prompt-based augmentation, we also conduct data augmentations on the original user historical behaviors in the prompt-enhanced sequences. Following previous CL-based models \cite{xie2022contrastive,zhou2020s3}, we randomly zero-mask proportional items in user behavior sequence with the mask ratio $\gamma_2$. Intuitively, the augmented prompt-enhanced sequence is noted as $\bar{s}_u^2=\{{p}_1^u,{p}_2^u, ...,{p}_n^u,v_1^u,[mask], ...,v_{|s_u|}^u\}$, strengthening the sequence modeling ability from another aspect. These two augmentations cooperate well in SSL.
        
        \subsubsection{Contrastive Learning}
        
        In prompt-oriented contrastive learning, we hope PPR can distinguish whether two user representations learned from (augmented) prompt-enhanced behavior sequences derive from the same user. To achieve this goal, we need to minimize the differences between the original and augmented sequences of the same users while maximizing the gaps between different users' representations.
        Specifically, for a batch $B$ with size $N$,  we apply the above two augmentations to each user $u$, and get the augmented sequences $\bar{s}_u^1$ and $\bar{s}_u^2$. We regard the original and the corresponding augmented user behavioral representations of $u$ as the positive pair $(\bm{u}_s,\bm{\bar{u}}_s)$. The rest augmented user representations $\bm{\bar{u}}'_s$ of other users $u'$ in the batch form the negative set $S^u_-$ of $u$. We use the cosine similarity $\mathrm{sim}(\cdot,\cdot)$ to measure the similarity. Formally, the final CL loss function $L_{CL}$ is formulated as:
        \begin{equation}
        \begin{aligned}
            L_{CL}= - \sum_{B} \sum_{u \in U^c}\log\frac{\exp(\mathrm{sim}(\bm{u}_s,\bm{\bar{u}}_s)/\tau)}
            {\sum_{u'\in {B}}\exp(\mathrm{sim}(\bm{u}_s,\bm{\bar{u}}'_s)/\tau)}.
        \end{aligned}
        \end{equation}
        Here, the augmented user representation $\bm{\bar{u}}_s$ has two forms $f_{seq}(\bar{s}_u^1|\Theta,\vartheta)$ or $f_{seq}(\bar{s}_u^2|\Theta,\vartheta)$. $\tau$ is the temperature hyper-parameter.
        The prompt-oriented contrastive learning loss is used as an auxiliary task of $L_p$ in prompt-tuning to train the personalized prompts fully. The overall loss $L_{all}$ is defined with the loss weight $\lambda$ as follows:
        \begin{equation}
        L_{all} = L_p+\lambda L_{CL}.
        \end{equation}
        
        \subsection{Model Designs and Complexity}
        
        \noindent
        \textbf{Model Designs.}
        In NLP, pre-training with prompt performs better than classical fine-tuning, especially in few-shot scenarios. Recently, pre-training has also shown its power in recommendation \cite{sun2019bert4rec,wu2022selective}. Hence, we attempt to use prompt-tuning to help pre-trained recommendation models to \textbf{\emph{more effectively and parameter-efficiently fit downstream tasks}} in recommendation.
        The advantages of prompts and using prompts in our tasks are listed in Sec. \ref{sec.introduction} and Sec. \ref{sec.method}.
        
        We adopt user profiles to build the personalized prompts in tuning, since user profiles can prompt the warmed model to give more prior knowledge \emph{related to the current user} in sequential modeling, and they are naturally available and widely-used in practical systems. We restrict PPR to merely utilize user profiles in the downstream tuning for universality and better user privacy protection in potential future applications. We do not tune the warmed part, since we want the proposed PPR to be suitable for different potential large-scale pre-trained recommendation models.
        Moreover, we design two versions PPR(light) and PPR(full). It is because different from NLP, the tokens (i.e., items) in recommendation are usually more than a million levels. Hence, the item embeddings appearing in the downstream tasks should be better  tuned (just like the verbalizer tuned in prompt-tuning of NLP). Nevertheless, PPR(light) still achieves good performance. We could choose PPR versions flexibly according to practical demands.
        
        \noindent
        \textbf{Complexity.}
        PPR(light) is parameter-efficient, since it only needs to tune and store a few parameters compared with fine-tuning (i.e., $\Theta$ is fixed). In our datasets, the numbers of parameters to be updated in PPR(light) are merely $0.3\%$, $0.25\%$, and $0.6\%$ of those in fine-tuning, respectively. PPR(full) is designed specifically for better representation learning of million-level items in recommendation. It has comparable updated parameters with better performances. Table \ref{tab:complexity} shows the details statistics. In both tuning and serving, Fine-tuning and PPR models have similar time costs.

     \subsection{Discussion}
        \textbf{Pre-traning model selection.}
     In this paper, we adopt SASRec as the pre-trained model. Throughout, SASRec is not custom-designed for recommendation pre-training. We select it for the following reasons: (1) SASRec is a classic sequential recommendation model, and its effectiveness has been extensively validated.  (2) The SASRec-like architecture is widely adopted as the based model for various recommendation pre-trained models \cite{zhou2020s3,hou2023learn,xie2022contrastive,hou2022towards}. Besides, our PPR is the pre-trained model  (3) In this paper, we aim to propose a framework to alleviate user cold start problems and other downstream tasks.  In recommendation,  considering the sparse behavior and high noise in the actions of cold users, recommendations for cold and warm users are often regarded as distinct tasks. However, in reality, the fundamental task is the next item prediction task. Previous works have already demonstrated that SASRec is sufficient to accomplish this task. Furthermore, cross-domain recommendations and user attribute prediction share similar characteristics. 
      
        Furthermore, we would like to emphasize that, in this paper, although SASRec itself was not custom-designed for pre-training, the SASRec model trained on the pre-training dataset serves as a pre-trained model. It only has a single pre-training task, namely, next-item prediction. It signifies our model's capability to predict what a user will likely prefer in the next time step, which also is the objective of most recommendation systems.

        \begin{table}[!htbp]
        \caption{Detailed statistics of tuned parameters of three datasets.}
        \label{tab:complexity}
        \center
        \begin{tabular}{l|ccc}
        \toprule
        Dataset &\ fine-tuning&\ PPR(light)  & PPR(full)\\
        \midrule
        \multirow{1}{*}{CIKM}
        ~ &100\%&0.3\%&100.3\%\\
        \multirow{1}{*}{QQBrowser}
        ~ &100\%&0.25\%&100.2\%\\
        \multirow{1}{*}{AliEC}
        ~ &100\%&0.6\%&100.6\%\\
        \bottomrule
        \end{tabular}
        \end{table}
        

        
        %
        %
        
        
        
        
        \section{Experiments}
        
        In this section, we conduct extensive experiments to answer the following six research questions:
        (\textbf{RQ1}): How does PPR perform in few-shot scenarios (Sec. \ref{sec.few-shot})?
        (\textbf{RQ2}): Can PPR work well in the challenging zero-shot recommendation (Sec. \ref{sec.zero-shot})?
        (\textbf{RQ3}): What are the effects of different components in PPR (Sec. \ref{sec.ablation})?
        (\textbf{RQ4}): Can PPR also work on different based models (Sec. \ref{sec.universality})?
        (\textbf{RQ5}): Can PPR achieve improvements under different data sparsity (Sec. \ref{sec.sparsity})?
        (\textbf{RQ6}): Is PPR still effective to be adopted on other downstream recommendation tasks (Sec. \ref{sec.explorations})?
        
        \subsection{Datasets}
        \label{sec.dataset}
        
        We evaluate PPR on three real-world open datasets, namely CIKM, QQBrowser, and AliEC\&AliAD. In all datasets, the users are split into warm users and cold-start users according to a threshold of interacted items (i.e., users having less than $10$ clicks are empirically viewed as cold-start users\cite{lee2019melu,ding2018improving}).
        In real-world systems, the behavioral patterns and recommendation models of cold-start users are usually different from those of warm users. Hence, we regard the general recommendation (warm-training) and the cold-start recommendation as two tasks.
        The click instances of warm users are used as the warm-training set, while those of cold-start users are used as the tuning/test set for downstream tasks. For cold-start user recommendation, we randomly split the cold-start users into train (in tuning) ($80\%$) and test ($20\%$) sets  following \cite{lee2019melu}. More details are in Table\ref{tab:dataset}.
        \begin{table}[!htbp]
        \caption{Detailed statistics of three datasets.}
        \label{tab:dataset}
        \center
        \begin{tabular}{l|cccc}
        \toprule
        Dataset &\# user&\# item&\tabincell{c}{\# warm-training\\ instance}& \tabincell{c}{\# tuning \\ instance}\\
        \midrule
        \multirow{1}{*}{CIKM}
        ~ &80,964&87,894&2,103,610&143,726\\
        \multirow{1}{*}{QQBrowser}
        ~ &134,931&97,904&15,359,880&263,572\\
        \multirow{1}{*}{AliEC}
        ~ &104,984&109,938&8,768,915&39,292\\
        \bottomrule
        \end{tabular}
        \end{table}
        
        \begin{table*}[!htbp]
        \caption{Results on few-shot recommendation. All improvements are significant over baselines (t-test with p ${<}$ 0.05).}
        \label{tab:few-shot}
        \center
        \begin{tabular}{l|l|ccccccccc}
        \toprule
        \small
        Dataset & Model &AUC& HIT@5& NDCG@5&  HIT@10& NDCG@10&HIT@20& NDCG@20&HIT@50& NDCG@50 \\
        \midrule
        \multirow{8}{*}{CIKM}
        ~ & BERT4Rec & 0.8482&0.5191&0.4279&0.6235&0.4616&0.7367&0.4902&0.8965&0.5220 \\
        ~ & SASRec & 0.8676 &0.5808 &0.4924 &0.6722 &0.5220 &0.7717 & 0.5471 &0.9100 & 0.5746\\
        ~ & MeLU & 0.8255 & 0.4647 & 0.3788 &0.5692 & 0.4124 &0.6963
        & 0.4444 &0.8773 &0.4803 \\
        ~ & CL4SRec & 0.8676& 0.5840&0.4930&0.6737&0.5252&0.7726& 0.5469&0.9099&0.5742\\
         ~ &   GCE-GNN & 0.8676&0.5156&0.3901&0.6529&0.4345&0.7847&0.4679&0.9240&0.4950\\
       ~&  PeterRec &0.8686&0.4950&0.3578&0.6471&0.4070&0.7852&0.4420&0.9290&0.4708\\   
        ~ &warm-train&0.8630&0.5779 & 0.4906 & 0.6695 & 0.5203 & 0.7660 & 0.5446&0.9039&0.5721\\
        ~ &fine-tuning&0.8731&0.5886&0.4948&0.6836&0.5255&0.7837&0.5508&0.9159&0.5772\\
        \cmidrule{2-11}
        ~ &PPR(light)&\textbf{0.8780}&0.5895&0.4955&0.6854&0.5265&0.7894&0.5528&\textbf{0.9230}&0.5795\\
        ~ &PPR(full)&0.8774&\textbf{0.5918}&\textbf{0.4976}&\textbf{0.6889}&\textbf{0.5290}&\textbf{0.7903}&\textbf{0.5547}&0.9211&\textbf{0.5807}\\
        \midrule
        \midrule
        \multirow{8}{*}{QQBrowser}
        ~ & BERT4Rec &0.9546&0.7782&0.6295&0.8734&0.6606&0.9386&0.6772&0.9858&0.6867 \\
        ~ & SASRec
        & 0.9629 &0.8077 &0.6609 &0.8974 &0.6902 & 0.9543 & 0.7048 &0.9904 &0.7121\\
        ~ & MeLU
        & 0.9318 & 0.6616 & 0.5020 & 0.7929 & 0.5447& 0.8981&0.5714 & 0.9810 & 0.5882 \\
        ~ & CL4SRec & 0.9626&0.8064&0.6543&0.8958&0.6864&0.9550&0.7015&0.9898&0.7078\\
        ~ & GCE-GNN&0.9592&0.7837&0.6258&0.8817&0.6577&0.9490&0.6749&0.9888&0.6830\\
       
         ~ & PeterRec & 0.9618&0.7860&0.6289&0.8883&0.6622&0.9560&0.6795&0.9900&0.6871\\
        ~ &warm-train&0.9572&0.7842&0.6338&0.8818&0.6657&0.9448&0.6817&0.9873&0.6904\\
        ~ &fine-tuning&0.9645&0.8142&0.6659&0.9017&0.6944&0.9578&0.7087&0.9919&0.7156\\
        \cmidrule{2-11}
        ~&PPR(light)&0.9640&0.8068&0.6545&0.8982&0.6843&0.9575&0.6994&\textbf{0.9926}&0.7066\\
        ~&PPR(full)&\textbf{0.9652}&\textbf{0.8169}&\textbf{0.6680}&\textbf{0.9031}&\textbf{0.6960}&	\textbf{0.9589}&	\textbf{0.7104}&0.9920&	\textbf{0.7171}\\
        \midrule
        \midrule
        \multirow{9}{*}{AliEC} 
        
        ~ & BERT4Rec & 0.8758&0.5543&0.4370&0.6738&0.4757&0.7898&0.505&0.9266&0.5324\\
        ~ & SASRec &
        0.8842 & 0.5884 & 0.4663 & 0.6997 &0.5024 &0.8083 &0.5299 &0.9316 &0.5546 \\
        ~ & GRU4Rec &0.8776 & 0.5974 & 0.4743 & 0.7009 & 0.5078 & 0.8002 & 0.5330 & 0.9192 & 0.5567 \\
        ~ & MeLU &0.8900&0.5759&0.4512&0.6955&0.4899&0.8131&0.5197&0.9326&0.5457\\
        
        ~ & CL4SRec & 0.8900&0.6036&0.4778&0.7137&0.5134&0.8239&0.5413&0.9379&0.5641\\
         ~ &  GCE-GNN & 0.8817&0.5746&0.4574&0.6932&0.4957&0.8102&0.5254&0.9299&0.5490\\
          ~ & PeterRec&0.8838&0.5395&0.3980&0.6792&0.4432&0.8087&0.4761&0.9414&0.5027\\    
      
        ~& warm-train& 0.8838&0.5880&0.4710&0.7006&0.5073&0.8110&0.5351&0.9308&0.5592\\
        ~&fine-tuing&0.8907&0.6058&0.4851&0.7189&0.5217&0.8212&0.5475&0.9371&0.5708\\
        \cmidrule{2-11}
        ~&PPR(light)&\textbf{0.8975}&0.6123&0.4873&\textbf{0.7275}&0.5246&\textbf{0.8363}&0.5521&\textbf{0.9422}&0.5734\\
        ~&PPR(full)&0.8941&\textbf{0.6126}&\textbf{0.4896}&0.7241&\textbf{0.5256}&0.8284&\textbf{0.5522}&0.9374&\textbf{0.5739}\\
        \bottomrule
        \end{tabular}
        \end{table*}

        \textbf{CIKM}. The CIKM dataset is an E-commerce recommendation dataset released by Alibaba\footnote{https://tianchi.aliyun.com/competition/entrance/231719\\/introduction}. It has $60$ thousand warm users with $2.1$ million click instances in the warm-train set. Other $21$ thousand cold-start users with $143$ thousand instances are used for tuning and testing. Each user has $3$ attributes: gender, age, and consumption level.
        
        \textbf{QQBrowser}. This dataset is collected from QQ Browser \cite{yuan2020parameter} on news/videos. It has $107$ thousand warm users and $28$ thousand cold users. Each user has $3$ attributes: gender, age, and life status.
        
        \textbf{AliEC\&AliAD}. This dataset contains two sub-datasets: AliEC for E-commerce and AliAD for advertising\footnote{https://tianchi.aliyun.com/dataset/dataDetail?dataId=56}. AliAD is much sparser than AliEC. For cold-start recommendation and user profile prediction tasks, we evaluate on AliEC. It has nearly $99$ thousand warm users and $6.4$ thousand cold users.  For cross-domain recommendation, we set AliEC as the source domain and AliAD as the target domain. Each user has $8$ attributes: user SEGID, user group ID, gender, age, consumption level, shopping level, occupation, and city.

        
        \subsection{Competitors}
        \label{sec.baseline}
        
        In this work, we adopt the representative SASRec \cite{kang2018self} as our base model, while it is also convenient to deploy PPR on other base recommendation models (e.g., CL4SRec \cite{xie2022contrastive}). In few-shot recommendation. We compare PPR with several competitive models as follows:
        
        (1) \textbf{SASRec} \cite{kang2018self}, which is a classical sequential recommendation model that introduces self-attention in behavior modeling. It is optimized by both warm-train and tuning sets.
        It is used as PPR's base model.

        (2) \textbf{BERT4Rec} \cite{sun2019bert4rec}, which is a widely-used pre-training recommendation model based on BERT. It uses masked item prediction as its pre-training task.
        
        (3) \textbf{MeLU} \cite{lee2019melu}, which is a typical meta-learning based cold-start recommendation model. It views new users as new tasks and adopts MAML for fast adaptations.
       
        (4) \textbf{CL4SRec} \cite{xie2022contrastive}, which is one of the SOTA CL-based sequential recommendation models also based on SASRec. It alleviates the sparsity and cold-start issues with several sequence-based augmentations and contrastive learning. We also deploy our PPR framework with CL4SRec in Sec. \ref{sec.universality}.

        (5) \textbf{GCE-GNN} \cite{wang2020global}, which is a session-based recommendation based on graph neural network. GCE-GNN converts user behavior sequence into a global and local graph, thus both considered local context information and global context information.
        
        (6) \textbf{PeterRec} \cite{yuan2020parameter}, which is a parameter-efficient model specifically designed for NextItNet \cite{yuan2019simple}. It fits downstream tasks through the proposed separate patches module.
        
        (7) \textbf{Warm-train}. We directly conduct the warmed model (SASRec) on the test set as a baseline for comparison.
        
        (8) \textbf{Fine-tuning}. It is a widely-used and strong tuning method to utilize warmed models for downstream tasks. It tunes all warmed model's parameters in tuning. Note that the fine-tuning model also jointly considers the same user profiles and historical behaviors as in Fig. \ref{fig:overall}.
        
        We also implement two PPR versions, namely PPR(light) and PPR(full) as introduced in Sec. \ref{sec.tuning_PPR}. PPR(light) only updates the parameters $\vartheta$ of the prompt generator and $\phi$ of the user profile learner with other parameters fixed, which is more parameter-efficient compared to fine-tuning for $\vartheta$ and $\phi$ is far smaller than the warmed model's parameters $\Theta$ (less than $1\%$ in Table \ref{tab:complexity}). PPR(full) updates all parameters as fine-tuning.
        
        We should highlight that all models share the \textbf{same input features} (including \emph{\textbf{user profiles}} encoded by the Deep model and all behaviors) and training instances in tuning for fair comparisons (except for warm-training that only uses the behavioral information of the warm-train set). Specifically, Same with our PPR, the user profiles first translate to corresponding embeddings and are inputted into the deep model, then the output of the deep model is plus with encoded user behavior. Besides, Fine-tuning and PPR share the same warmed model.
        
        \subsection{Experimental Settings}
        \label{sec.experimental_settings}
        
        \textbf{Parameter Settings.}
        For fair comparisons, the embedding size is $64$ and batch size is $256$ equally for all methods. We optimize all models by Adam. We conduct a grid search for hyper-parameters. We search models' learning rates among $\{1e-3, 3e-4,1e-4,3e-5,1e-5,3e-6,1e-6\}$. The best learning rates are $1e-5$ for fine-tuning and PPR(full), and $1e-4$ for the rest models.
        In PPR, the prompts are generated via Eq. (\ref{eq.prompt}) taking user profiles as input features, where $\mathrm{MLP}$ is verified strong enough. For efficiency, the prompt length is set as $1$ (we find that longer prompts do not bring in much improvement). The mask ratios of two data augmentations are $0.2$ and the loss weight $\lambda=0.1$.
        \begin{table*}[!h]
        \caption{Results on zero-shot recommendation. The improvements are significant over baselines (t-test with p$<$0.05).}
        \label{tab:zero-shot}
        \center
        \begin{tabular}{l|l|ccccccccc}
        \toprule
        \small
        Dateset & Model &AUC& HIT@5& NDCG@5&  HIT@10&NDCG@10&HIT@20&NDCG@20&HIT@50& NDCG@50 \\
        \midrule
        \multirow{3}{*}{CIKM}
        ~&fine-tuning&0.7753&0.2991&0.2061&0.4235&0.2463&0.5946&0.2894&0.8460&0.3394\\
        ~&PPR(light)&\textbf{0.7845}&0.2998&0.2063&\textbf{0.4337}&0.2493&\textbf{0.6041}&\textbf{0.2923}&\textbf{0.8621}&0.3436\\
        ~&PPR(full)&0.7825&\textbf{0.3069}&\textbf{0.2100}&0.4329&\textbf{0.2507}&0.5974&0.2920&0.8616&\textbf{0.3446}\\
        \midrule
        \multirow{3}{*}{QQBrowser}
        ~ &fine-tuning&0.8743&0.4950&0.3538&0.6427&0.4016&\textbf{0.7822}&0.4368&0.9393&0.4684\\
        ~&PPR(light)&0.8635&0.4829&0.3418&0.6337&0.3907&0.7770&0.4269&0.9262&0.4569\\
        ~&PPR(full)&\textbf{0.8757}&\textbf{0.5012}&\textbf{0.3612}&\textbf{0.6443}&\textbf{0.4074}&0.7813&\textbf{0.4421}&\textbf{0.9411}&\textbf{0.4742}\\
        \midrule
        \multirow{3}{*}{AliEC}
        ~&fine-tuning&0.8377&0.3905&0.2713&0.5675&\textbf{0.3286}&0.7252&0.3686&0.9006&\textbf{0.4039}\\
        ~&PPR(light)&0.8349&\textbf{0.4014}&\textbf{0.2743}&0.5575&0.3247&\textbf{0.7337}&\textbf{0.3693}&0.9022&0.4028\\
        ~&PPR(full)&\textbf{0.8380}&0.3983&0.2708&\textbf{0.5691}&0.3256&0.7283&0.3661&\textbf{0.9037}&0.4013\\
        \bottomrule
        \end{tabular}
        \end{table*}
        
        \noindent
        \textbf{Evaluation Protocols.}
        We mainly focus on the few-shot and zero-shot recommendation tasks of cold-start users.
        For all cold-start users in the test set, their first clicked items are used for zero-shot recommendation (since there is no historical behavior at this time), while the rest click behaviors are used for few-shot recommendation.
        We use the classical AUC, top-N hit rate (HIT@N), and Normalized Discounted Cumulative Gain (NDCG@N) as our evaluation metrics. For HIT@N and NDCG@N, we report top $5$, $10$, $20$ and $50$. For each ground truth, we randomly sample $99$ items that the user did not click as negative samples \cite{zhou2020s3,sun2019bert4rec}.


        \subsection{Few-shot Recommendation (RQ1)}
        \label{sec.few-shot}
        
        We first evaluate models on the few-shot recommendation. Table \ref{tab:few-shot} shows the results on three datasets. We can find that:
        
        (1) Our personalized prompt-based model outperforms all baselines on all metrics in three datasets (the significance level is $p<0.05$ via t-test). It indicates that our personalized prompts can better extract useful information related to the current user from the huge knowledgeable pre-training models, which is beneficial, especially for cold-start scenarios.
        We should highlight that fine-tuning is a very strong baseline (some prompt-tuning models in NLP only have comparable or even worse performance compared to fine-tuning). Nevertheless, PPR models still outperform fine-tuning and other classical cold-start recommendation models, including MeLU and CL4SRec (note that these baselines also have the same \emph{user profiles} as inputs). The advantages of PPR mainly derive from that (a) PPR conducts a unified well-learned sequential modeling on both user profiles and few-shot behaviors, and (b) user profiles are considered to enable a personalized user behavior modeling. Both factors are beneficial, especially in cold-start scenarios.
        
        (2) Comparing among different PPR settings, we observe that PPR(full) generally achieves better results in few-shot scenarios. Different from NLP tasks such as sentiment analysis with several labels, the label set of recommendation is far larger (often million-level). Moreover, items in recommendation do not have concrete semantics as words. Hence, the current warmed recommendation models cannot perfectly learn all item representations, and tuning is still beneficial. It is natural that the tuned item representations of PPR(full) can further improve the performances with comparable time and memory costs of fine-tuning.
        
        (3) PPR(light) achieves some SOTA results on HIT@N in AliEC dataset, and generally outperforms fine-tuning in two datasets. It is challenging since PPR(light) only tunes the personalized prompt part with all warmed model's parameters unchanged. The effectiveness and efficiency improvements of PPR(light) are encouraging as other prompts in NLP \cite{li2021prefix,liu2022p}. PPR(light) outperforms PPR(full) in a sparser AliEC dataset, which confirms that it is difficult to tune all warmed model's parameters with extremely sparse downstream instances. It emphasizes the advantages of PPR(light) compared to fine-tuning and PPR(full) besides efficiency. We can select between the full and light PPR versions according to the practical priority of effectiveness and efficiency.
        
        (4) Our PPR still outperforms the parameter-efficient model PterRec, it shows the power of our PPR.  Moreover, PterRec is a meticulously crafted pre-trained model with carefully designed parameter-efficient tuning. However, PterRec's parameter-efficient tuning is closely linked to its specific pre-trained model, making it model-specific. Our PPR is model-agnostic, which makes our PPR more flexible.
        
        \subsection{Zero-shot Recommendation (RQ2)}
        \label{sec.zero-shot}
        
        The data sparsity issue is extremely serious, where zero-shot users widely exist in practical systems. Most conventional sequential recommendation models (including SASRec, MeLU, and CL4SRec) cannot handle the zero-shot scenarios, since there is no historical behavior. We attempt to jointly address the zero-shot recommendation with the same PPR framework.
        For PPR, we directly input user profiles into the prompt generator as the few-shot recommendation without behavioral inputs. For fine-tuning baseline, only the user profile learner (i.e., the MLP in Eq. (\ref{eq.prompt_user})) is activated to learn from user profiles. From Table \ref{tab:zero-shot} we can observe that:
        
        \begin{table*}[htbp]
        \caption{Results of joint cold-start recommendation (few-shot+zero-shot). Improvements are significant (t-test with p$<$0.05).}
        \label{tab:joint_few_zero}
        \center
        \begin{tabular}{l|l|ccccccccc}
        \toprule
        \small
        Database & Model &AUC& HIT@5& NDCG@5&  HIT@10&NDCG@10&HIT@20&NDCG@20&HIT@50& NDCG@50 \\
        \midrule
        \multirow{3}{*}{CIKM}
        ~ &fine-tuning&0.8418&0.5202&0.4329&0.6163&0.4639&0.7257&0.4915&0.8891&0.5240\\
        ~&PPR(light)&\textbf{0.8606}&\textbf{0.5399}&0.4445&\textbf{0.6427}&0.4777&\textbf{0.7562}&0.5064&\textbf{0.9110}&0.5372\\
        ~&PPR(full)&0.8585&0.5396&\textbf{0.4476}&0.6409&\textbf{0.4802}&0.7522&\textbf{0.5084}&0.9068&\textbf{0.5392}\\
        \midrule
        \multirow{3}{*}{QQBrowser}
        ~ &fine-tuning&0.9593&0.7940&0.6459&0.8838&0.6751&0.9469&0.6912&0.9899&0.6999\\
        ~&PPR(light)&0.9589&0.7866&0.6350&0.8812&0.6658&0.9471&0.6826&\textbf{0.9904}&0.6914\\
        ~&PPR(full)&\textbf{0.9602}&\textbf{0.7965}&\textbf{0.6490}&\textbf{0.8863}&\textbf{0.6782}&\textbf{0.9486}&\textbf{0.6941}&0.9902&\textbf{0.7026}\\
        \midrule
        \multirow{3}{*}{AliEC}
        ~ &fine-tuning&0.8818&0.5708&0.4480&0.6912&0.4869&0.8057&0.5158&0.9317&0.5411\\
        ~&PPR(light)&\textbf{0.8904}&\textbf{0.5729}&0.4465&\textbf{0.7032}&0.4886&\textbf{0.8173}&0.5177&\textbf{0.9426}&0.5428\\
        ~&PPR(full)&0.8817&0.5726&\textbf{0.4519}&0.6934&\textbf{0.4908}&0.8050&\textbf{0.5190}&0.9308&\textbf{0.5443}\\
        \bottomrule
        \end{tabular}
        \end{table*}
       
        (1) Our PPR models still achieve the best performances on most metrics of three datasets, which confirms the effectiveness of PPR in zero-shot user understanding via fully reusing pre-training knowledge. We should highlight that getting improvements in zero-shot scenarios is challenging, since all models share the same sparse training instances and user profiles in tuning. Compared with fine-tuning, PPR makes full use of warmed models to encode user profiles, which is the main reason for our improvements.
        
        (2) Currently, the user profiles for prompt construction are not that sufficient in the open datasets, which may limit the modeling ability of prompt-tuning. It can be expected that the improvements will be more significant if enhanced with more user features. We have also evaluated the effectiveness of other prompt input features and construction methods (e.g., built via user behavioral embeddings learned from other domains) for other tasks in Sec. \ref{sec.explorations}.
        
        \textbf{Joint Few-shot and Zero-shot Recommendation.}
        Table \ref{tab:few-shot} and \ref{tab:zero-shot} demonstrate the effectiveness of our PPR in both few-shot and zero-shot recommendation. Considering the storage efficiency and maintenance cost in online deployment, we further conduct a challenging evaluation, using one set of model parameters to jointly address two cold-start recommendation tasks.
        Table \ref{tab:joint_few_zero} shows the results of these  joint cold-start recommendation task. We find that: PPR models significantly outperform fine-tuning by a larger margin in this joint scenarios. It indicates that the powerful personalized prompts could help to understand both few-shot and zero-shot users simultaneously in practice, which cannot be accomplished well by most classical sequential models as long as they rely on user behaviors.


        
        \subsection{Ablation Study (RQ3)}
        \label{sec.ablation}
        \begin{figure*}[!htbp]
        \centering
        \includegraphics[width=0.95\textwidth]{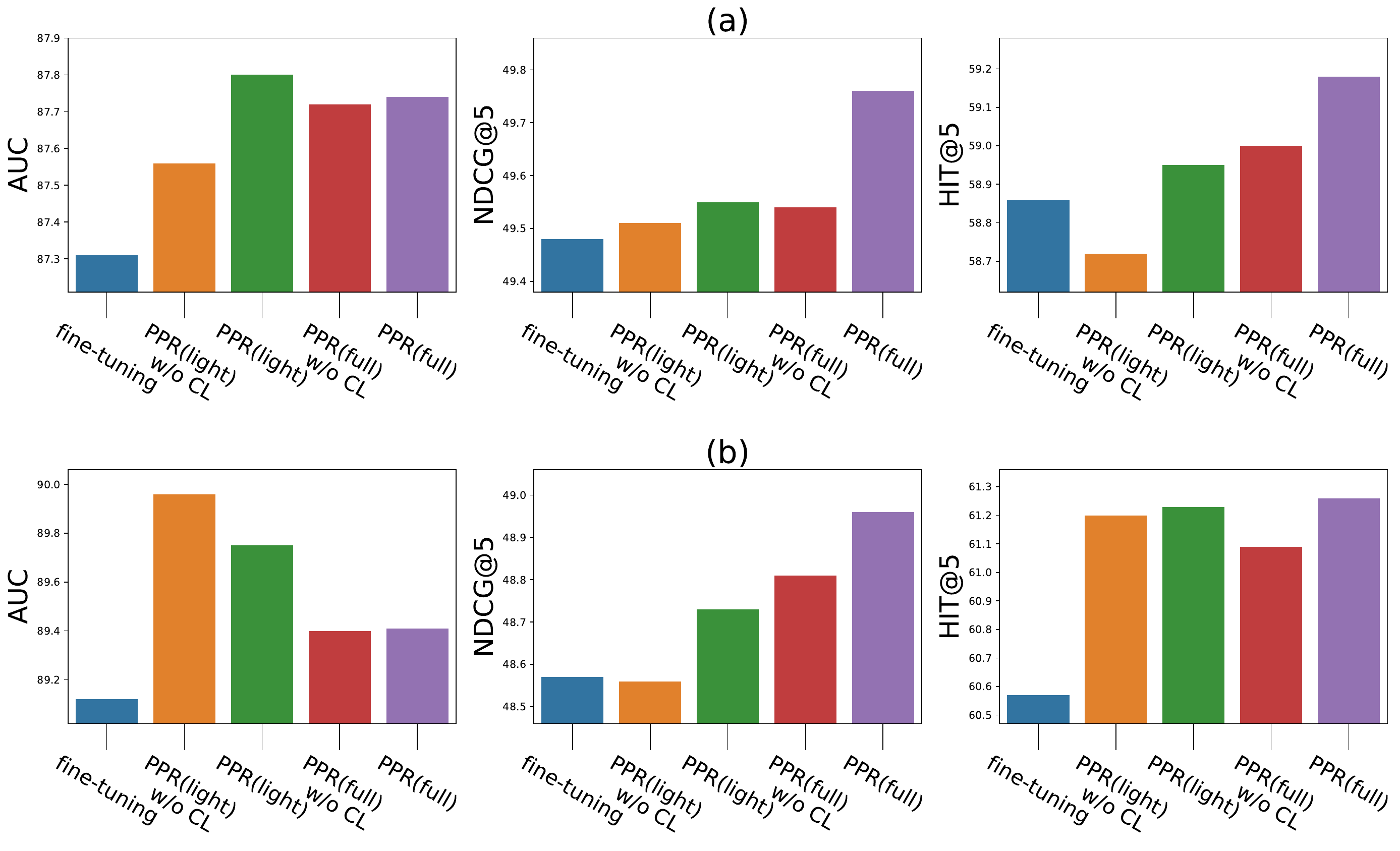}
        \caption{Ablation study of PPR(light) and PPR(full) on (a) CIKM, and (b) AliEC.All improvements are significant over baselines (t-test with p ${<}$ 0.05).}
        \label{fig:ablation_study}
        \end{figure*}

        In this section, we aim to confirm that the prompt-oriented contrastive learning is essential for PPR. Fig. \ref{fig:ablation_study} shows the results of different ablation versions of PPR(light) and PPR(full) on few-shot recommendation. We find that:
        
        (1) Generally, the prompt-oriented CL brings consistent improvements on almost all metrics on both datasets except for AUC on PPR(light). It confirms the effectiveness of our prompt-oriented CL with two types of augmentations for better prompt-enhanced sequence modeling.
        
        (2) PPR still outperforms fine-tuning on most metrics, even without the prompt-oriented CL. It verifies that our personalized prompt-tuning is truly superior to fine-tuning in transferring pre-trained knowledge to downstream tasks. Besides, we deploy PPR on a CL-based model CL4SRec as the base model and still achieve consistent improvements (in Sec. \ref{sec.universality}).

\begin{figure*}[!htbp]
        \centering
        \includegraphics[width=0.95\textwidth]{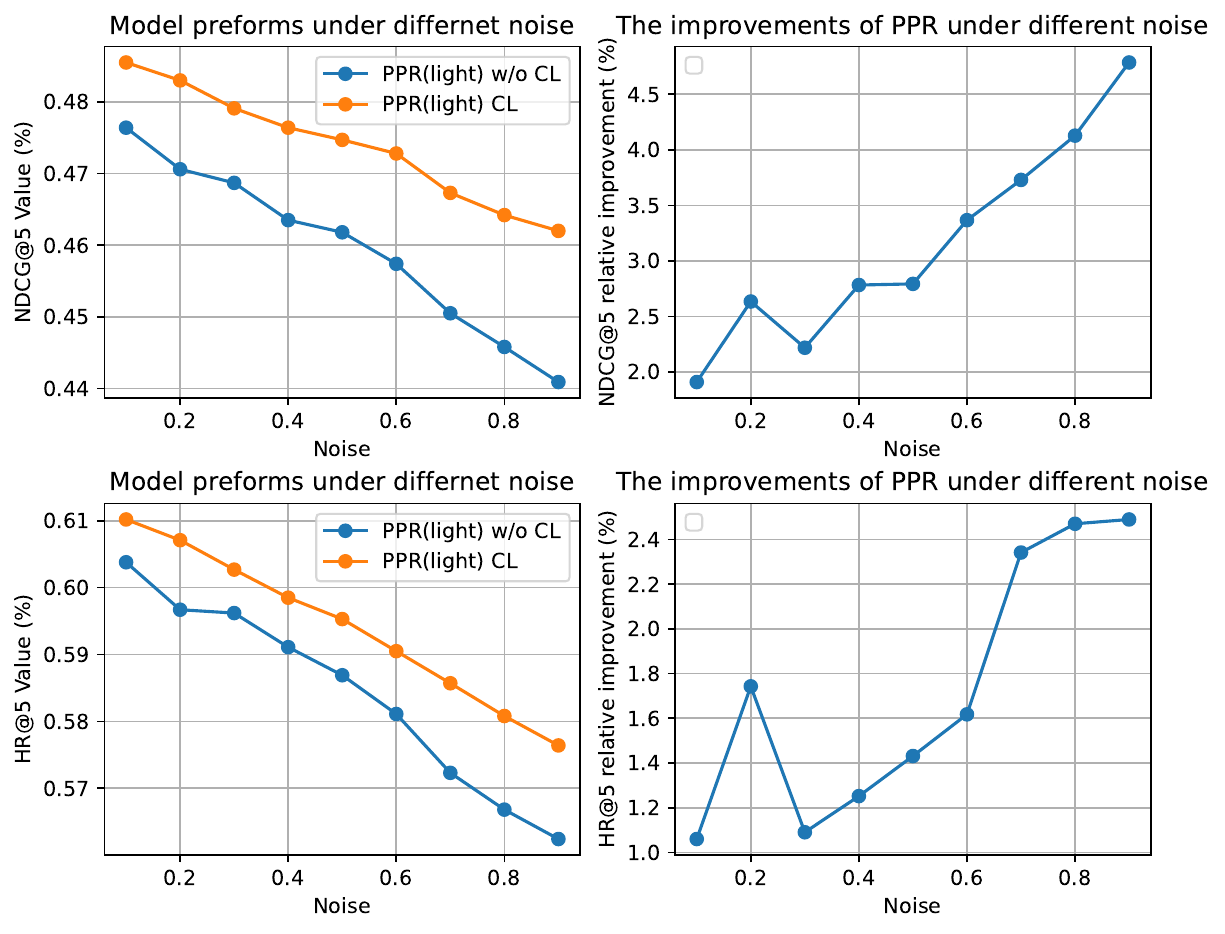}
        \caption{model performs under different noise on AliEC dataset}
        \label{fig:robustness evaluation of Contrastive learning }
\end{figure*}

        As discussed in Sec. \ref{sec.data_augmentation}, the cold user's user profile and behavior is noisy. Incorporating our contrastive learning task can make our PPR more robust. To evaluate it, we intentionally introduce noise to the user profiles. Specifically, for a user's profile $a_i^{u}$, we randomly replace it with another value $\hat{a}_i^{u}$  with the rate $r$. We change the noise rate from 0.1 to 0.9. We report the results in Fig. \ref{fig:robustness evaluation of Contrastive learning }. From it, We can observe that as noise increases, the performance of our model decreases. This phenomenon is natural since our PPR utilizes user profiles to generate prompts. Furthermore, our PPR consistently outperforms PPR w/o CL, and the gap between them continues to widen. This indicates with the help of our contrastive learning, our PPR can be more robust in noise data. In fact, in the real world, user data often contains noise, especially on smaller platforms that have limited user data. Our PPR may perform even better under those scenarios. 

        \subsection{Universality of PPR (RQ4)}
        \label{sec.universality}
        \begin{table*}[htbp]
        \caption{Results of PPR on few-shot recommendation based on CL4SRec. Improvements are significant (t-test with p$<$0.05).}
        \label{tab:few-shot on CL4SRec}
        \center
        \begin{tabular}{l|l|ccccccccc}
        \toprule
        \small
        Dataset & Model &AUC& HIT@5& NDCG@5&  HIT@10& NDCG@10&HIT@20& NDCG@20&HIT@50& NDCG@50 \\
        \midrule
        \multirow{4}{*}{CIKM}
        ~ &warm-train& 0.8590&0.5865&0.4942&0.6699&0.5252&0.7612&0.5481&0.8960&0.5750\\
        ~ &fine-tuning&0.8660&0.5948&0.5035&0.6797&0.5309&0.7719&0.5542&0.9034&0.5804\\
        \cmidrule{2-11}
        ~ &PPR(light)&0.8723&0.6013&0.5053&0.6891&0.5338&0.7851&0.5580&0.9106&0.5830\\
        ~ &PPR(full)&\textbf{0.8769}&\textbf{0.6066}&\textbf{0.5093}&\textbf{0.6982}&\textbf{0.5390}&\textbf{0.7928}&\textbf{0.5628}&\textbf{0.9157}&\textbf{0.5873}\\
        \midrule
        \midrule
        \multirow{4}{*}{QQBrowser}
        ~ &warm-train& 0.9554&0.7817&0.6304&0.8779&0.6618&0.9426&0.6783&0.9863&0.6872\\
        ~ &fine-tuning&0.9623&0.8083&0.6593&0.8971&0.6883&0.9548&0.7030&0.9896&0.7101\\
        \cmidrule{2-11}
        ~&PPR(light)&0.9641&0.8077&0.6566&0.8989&0.6863&\textbf{0.9583}&0.7015&\textbf{0.9919}&0.7084\\
        ~&PPR(full)&\textbf{0.9642}&\textbf{0.8138}&\textbf{0.6652}&\textbf{0.9008}&\textbf{0.6936}&0.9580&\textbf{0.7082}&0.9911&\textbf{0.7150}\\
        \midrule
        \midrule
        \multirow{4}{*}{AliEC}
        ~& warm-train& 0.8787&0.5878&0.4696&0.6977&0.5052&0.8021&0.5316&0.9248&0.5562\\
        ~&fine-tuning&0.8912&0.6057&0.4857&0.7226&0.5234&0.8242&0.5491&0.9365&0.5716\\
        \cmidrule{2-11}
        ~&PPR(light)&\textbf{0.8992}&0.6114&0.4857&\textbf{0.7328}&0.5250&\textbf{0.8386}&0.5518&\textbf{0.9441}&0.5729\\
        ~&PPR(full)&0.8934&\textbf{0.6140}&\textbf{0.4918}&0.7277&\textbf{0.5284}&0.8290&\textbf{0.5542}&0.9362&\textbf{0.5757}\\
        \bottomrule
        \end{tabular}
        \end{table*}
        PPR is an effective and universal tuning framework, which can be easily deployed on different base models. In this section, we further adopt PPR with CL4SRec \cite{zhou2020s3} used as the base model. The results of PPR models on few-shot recommendation based on CL4SRec are given in Table \ref{tab:few-shot on CL4SRec}, from which we can find that:
        
        (1) PPR models still achieve the best performances on all metrics in three datasets, which proves the universality of PPR with different base recommendation models. It implies that our PPR could function as a general enhanced tuning mechanism of fine-tuning when adopting large-scale pre-trained recommendation models for downstream tasks.
        
        (2) Generally, PPR(full) outperforms PPR(light) on most metrics. It indicates that tuning item embeddings (i.e., the labels to be predicted) is essential in cold-start recommendation, which differs from the prompt-tuning in NLP. Nevertheless, PPR(light) still performs better than fine-tuning on most metrics. Similar to Table \ref{tab:few-shot}, PPR(light) also has several better results in HIT@N. We can flexibly choose different implementations of PPR to enhance fine-tuning considering the practical demand on parameter-efficiency.

        \subsection{Model Analyses on Sparsity (RQ5)}
        \label{sec.sparsity}
        \begin{figure*}[!hbtp]
        \centering
        \includegraphics[width=0.95\textwidth]{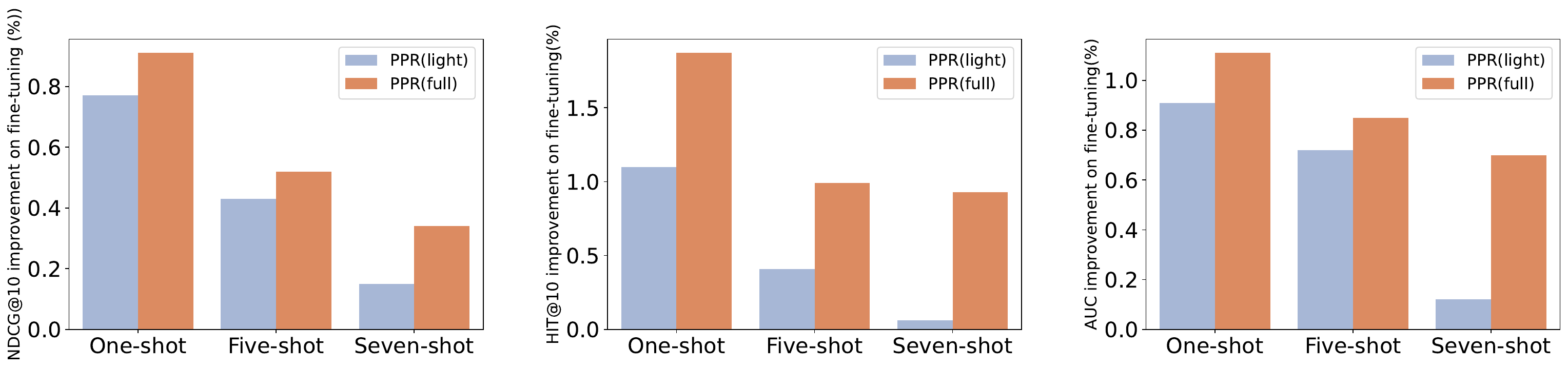}
        \caption{Relative improvements of PPR models over fine-tuning with different degrees of data sparsity in CIKM dataset.All improvements are significant over baselines (t-test with p ${<}$ 0.05).}
        \label{fig:Sparisty}
        \end{figure*}
        We further explore the influence of data sparsity on PPR to show its robustness. Specifically, we crop all user behavior sequences of both train and test sets in tuning to represent different degrees of sparsity. We define the k-shot setting where all user behavior sequences are cropped and no longer than $k$, with $k=1,5,7$. Fig. \ref{fig:Sparisty} displays the relative improvements of PPR(light) and PPR(full) over fine-tuning on AUC, HIT, and NDCG in CIKM. We have:
        
        (1) Both PPR(light) and PPR(full) consistently outperform fine-tuning with different sparsity, which verifies the robustness of PPR in different cold-start scenarios.
        
        (2) The improvements of PPR models increase with the maximum user behavior sequence length decreasing. It indicates that PPR can perform better on more sparse scenarios compared to fine-tuning. It is intuitive since the personalized prompts are more dominating when the behavioral information is sparser. In this case, the advantages of PPR (e.g., the better usage of warmed sequential modeling on prompt-enhanced sequence) are amplified.
        
        \subsection{Explorations of PPR on Other Tasks (RQ6)}
        \label{sec.explorations}
        
        Currently, pre-trained language models with prompt-tuning is dominating and widely verified as a paradigm in almost all NLP downstream tasks. Inspired by this, we believe that pre-trained recommendation model may also be a promising future direction for different recommendation tasks. Therefore, besides the cold-start recommendation task, we further explore other promising usages of personalized prompts on cross-domain recommendation and user profile prediction as an appropriate alternate of fine-tuning.
        Note that we \textbf{do not} want to rashly verify that PPR could achieve SOTA results on all recommendation tasks compared to well-designed sophisticated models. Instead, our explorations on other downstream tasks aim to shed light on the promising future direction of pre-training and prompt-tuning in recommendation.
        
        \subsubsection{PPR on Cross-domain Recommendation}
        \begin{table*}[htbp]
        \label{sec.cross-domain}
        \caption{Results of cross-domain recommendation on AliEC$\rightarrow$AliAD. All improvements are significant (t-test with p$<$0.01).}
        \label{tab:cross-domain}
        \center
        \begin{tabular}{l|l|ccccccccc}
        \toprule
        \small
        Dataset & Model &AUC& HIT@5& NDCG@5&  HIT@10&NDCG@10&HIT@20&NDCG@20&HIT@50& NDCG@50 \\
        \midrule
        \multirow{2}{*}{\tabincell{c}{AliEC \\$\rightarrow$ \\AliAD}}
        ~&SASRec(target)&0.6423&0.2752&0.2106&0.3590&0.2375&0.4594&0.2628&0.6568&0.3017\\
        ~&MiNet&0.6474&0.2573&0.1872&0.3499&0.2170&0.4507&0.2424&0.6561&0.2827\\
        ~&fine-tuning&0.7165&0.3362&0.2494&0.4357&0.2813&0.5552&0.3115&0.7323&0.3465\\
        ~&PPR(full)&\textbf{0.7343}&\textbf{0.3632}&\textbf{0.2799}&\textbf{0.4595}&\textbf{0.3109}&\textbf{0.5761}&\textbf{0.3402}&\textbf{0.7631}&\textbf{0.3772}\\
        \bottomrule
        \end{tabular}
        \end{table*}
        Cross-domain recommendation (CDR) aims to transfer useful knowledge from the source domain to help the target domain \cite{hu2018conet,ouyang2020minet}. We focus on the CDR scenario with overlapping users and non-overlapping items. In this work, the source domain (warm-train set) is AliEC and the target domain (tuning set) is AliAD. For CDR, we directly use the user embeddings trained on the source domain by the same warmed model as our personalized prompts. Similarly, these source-domain user embeddings are also applied as side information for fine-tuning and PPR. We also implement (a) SASRec(target), which is only trained on the target domain without source information, and (b) MiNet \cite{ouyang2020minet}, which is a classical CDR method, for comparisons. Note that there 
        are no overlapping items, which means we have to tune the new target-domain item embeddings. Hence, we choose the PPR(full) version. The results are shown in Table \ref{tab:cross-domain}.
        
        We can find that PPR significantly outperforms all baselines on all metrics. The only difference between PPR and fine-tuning is the personalized prompt (all models use the same source/target domain features). It implies the promising application of PPR on other downstream tasks where items of pre-training and tuning stages are even different.
        
        
        \subsubsection{PPR on User Profile Prediction}
        
        User profile prediction task aims to predict users' profiles via their behavior sequences \cite{xiao2021uprec}.
        We conduct an exploration of PPR on this task by predicting users' graduate state in AliEC, which is a binary classification task. Precisely in PPR, the personalized prompts are generated by users' all profiles except the one to be predicted. For classification, we add a profile classifier after the final user representation for both PPR and fine-tuning. The results are shown in Table \ref{tab:user_profile}.
        
        We can observe that PPR significantly outperforms fine-tuning on ACC, precision, and F1. It is  because that PPR can better interact users' profile information with their behavioral information via a unified warmed sequential model. We also evaluate PPR on other user profile prediction tasks, such as gender and age, where PPR(light) has comparable results with far less parameter tuning. In the future, we can explore designing  more customized prompts for different downstream tasks to improve their performances.
        
        \begin{table}[!htbp]
        \label{sec.user_profile}
        \caption{Results of user profile prediction on AliEC.All improvements are significant (t-test with p$<$0.05).}
        \label{tab:user_profile}
        \center
        \begin{tabular}{l|l|cccp{0.9cm}<{\centering}}
        \toprule
        \small
        Dataset & Model &ACC&Precision&Recall&F1 \\
        \midrule
        \multirow{2}{*}{AliEC}
        ~ &fine-tuning& 0.89&0.33&\textbf{0.82}&0.47 \\
        ~&PPR(light)&\textbf{0.92}&\textbf{0.42}&0.74&\textbf{0.54}\\
        \bottomrule
        \end{tabular}
        \end{table}
        
        
        
        

        
        
        
        \section{Conclusion and Future Work}
        
        In this work, we propose a new personalized prompt-based recommendation to improve the tuning of pre-trained recommendation models. We conduct extensive experiments to verify the effectiveness, robustness, and universality of our PPR on cold-start recommendation tasks, and also explore the potential wide usage of PPR on other downstream tasks.
        PPR is the first small step on adopting personalized prompts instead of fine-tuning in pre-trained recommendation models, shedding lights on future explorations.
        
        In the future, we will verify our PPR on more recommendation tasks with customized prompt-tuning manners, prompt generators and pre-training tasks. We will also explore the effectiveness of PPR on industrial extremely large-scale recommendation datasets and pre-trained models. Jointly considering pre-trained recommendation and language models also deserves further exploration.

        \section*{Acknowledgment}
        The research work is supported by the National Natural Science Foundation of China under Grant (No.61976204, No.62176014), the National Key Research and Development Program of China under Grant No. 2021ZD0113602, the Fundamental Research Funds for the Central Universities and Tencent.
        \addcontentsline{toc}{section}{Acknowledgment}

        \bibliographystyle{IEEEtran}
        \bibliography{reference}
        
        
        
        \begin{IEEEbiography}[{\includegraphics[width=1in,height=1.25in,clip,keepaspectratio]{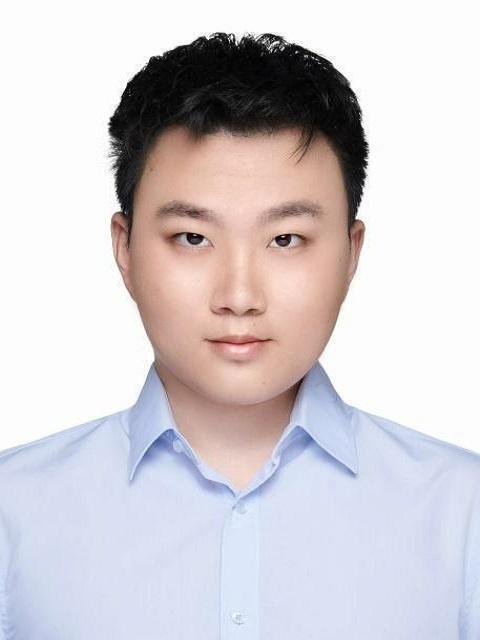}}]{Yiqing Wu}
        is currently pursuing his M.S. degree in the Institute of Computing Technology, Chinese Academy of Sciences, Beijing, China.  He recieved his B.E. degree from Dalian University of Technology in 2020. His main research interests include recommender system, graph neurnal network and data mining.
        \end{IEEEbiography}
        \begin{IEEEbiography}[{\includegraphics[width=1in,height=1.25in,clip,keepaspectratio]{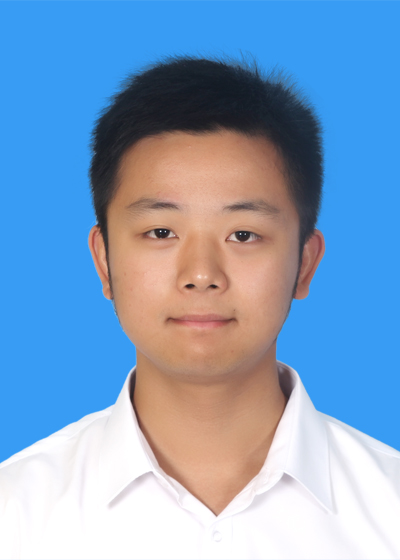}}]{Ruobing Xie} is a senior researcher of WeChat, Tencent. He received his BEng degree in 2014 and his master degree in 2017 from the Department of Computer Science and Technology, Tsinghua University. His research interests include recommender system, knowledge graph, and natural language processing. He has published over 60 papers in top-tier conferences and journals including KDD, WWW, SIGIR, ACL, AAAI, NeurIPS and TKDE. He is a member of CIPS Youth Working Committee and Social Media Processing Committee.
        \end{IEEEbiography}

        \begin{IEEEbiography}[{\includegraphics[width=1in,height=1.25in,clip,keepaspectratio]{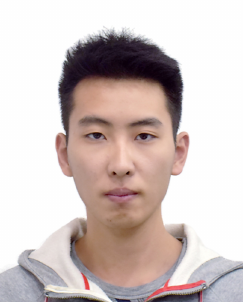}}]{Yongchun Zhu}
        is currently pursuing his Ph.D. degree in the Institute of Computing Technology, Chinese Academy of Sciences, Beijing, China. His main research interests include transfer learning, fake news detection and recommender system. He has published over 20 papers in journals and conference proceedings including KDD, WWW, SIGIR, TNNLS and so on.
        \end{IEEEbiography}
        
        \begin{IEEEbiography}[{\includegraphics[width=1in,height=1.25in,clip,keepaspectratio]{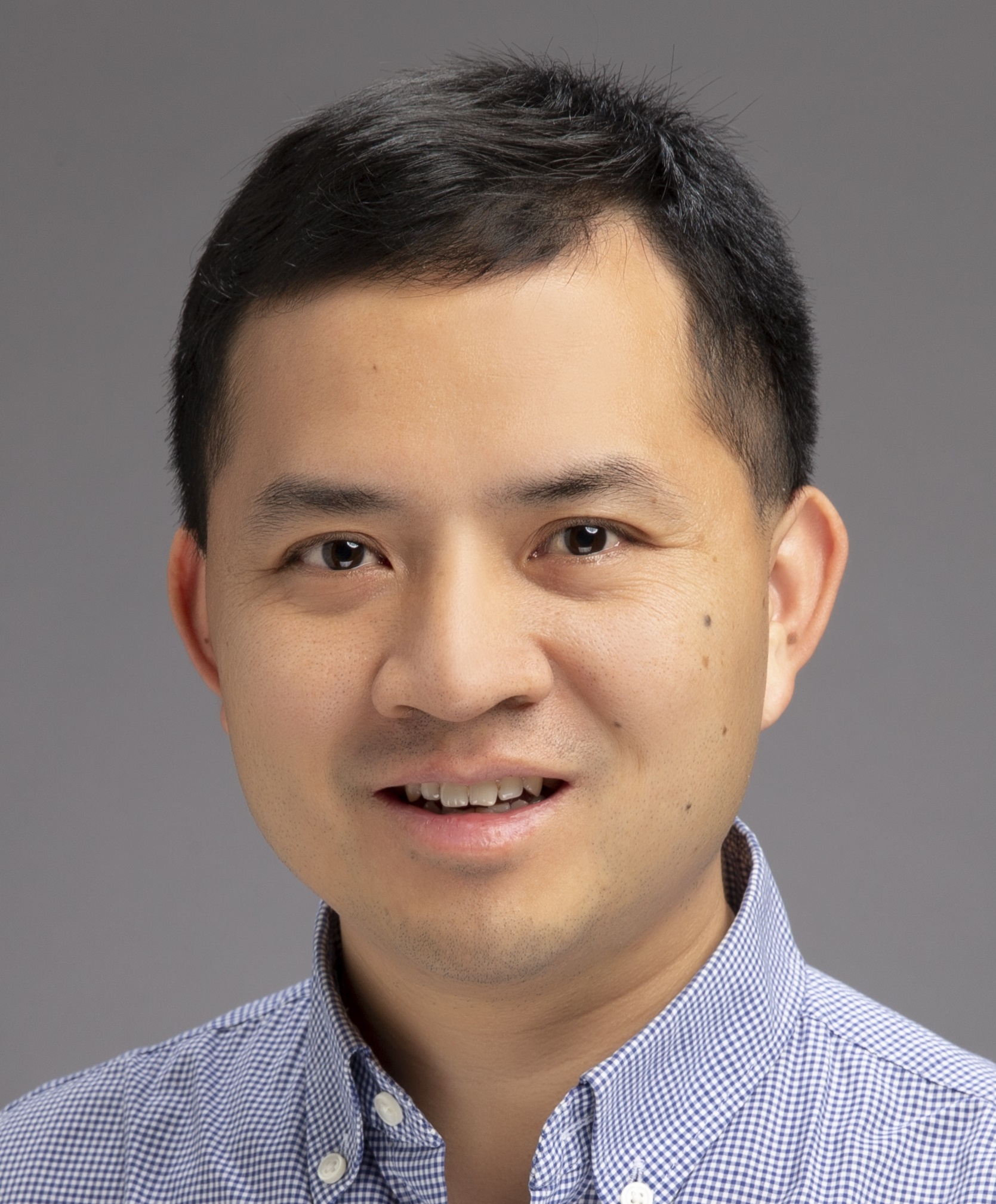}}]{Fuzhen Zhuang}
        is a professor in Institute of Artificial Intelligence, Beihang University. His research interests include transfer learning, machine learning, data mining, multi-task learning and recommendation systems. He has published over 100 papers in the prestigious refereed journals and conference proceedings, such as Nature Communications, TKDE, Proc. of IEEE, TNNLS, TIST, KDD, WWW, SIGIR, NeurIPS, AAAI, and ICDE. 
        \end{IEEEbiography}
        
        \begin{IEEEbiography}[{\includegraphics[width=1in,height=1.25in,clip,keepaspectratio]{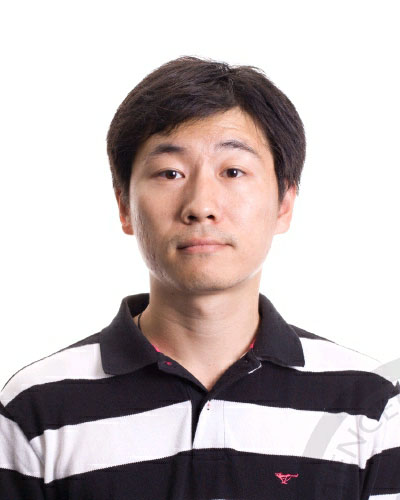}}]{Xu Zhang} received the master’s degree in College of Computer Science and Engineering, Northeastern University in 2008. He is the team leader of WeChat User Profile, Tencent. His research interests include machine learning and its applications, such as social network, user interest model and user behavior research. He has published more than 10 papers in top conferences such as KDD, SIGIR, CIKM.
        
        \end{IEEEbiography}
        
        \begin{IEEEbiography}[{\includegraphics[width=1in,height=1.25in,clip,keepaspectratio]{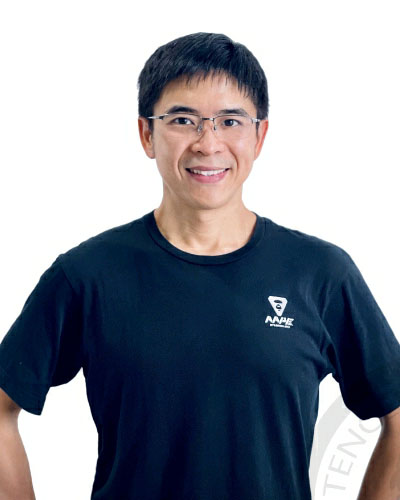}}] {Leyu Lin} received the masters degree in Institute of Computing Technology, Chinese Academy of Sciences, in 2008. He is currently the director of WeChat Recommendation Product Center, Tencent. His research interests include information retrieval and its applications, such as search system, recommendation system and computational advertising. He has published more than 20 papers in top conferences such as KDD, SIGIR, WWW.

        \end{IEEEbiography}
        
        \begin{IEEEbiography}[{\includegraphics[width=1in,height=1.25in,clip,keepaspectratio]{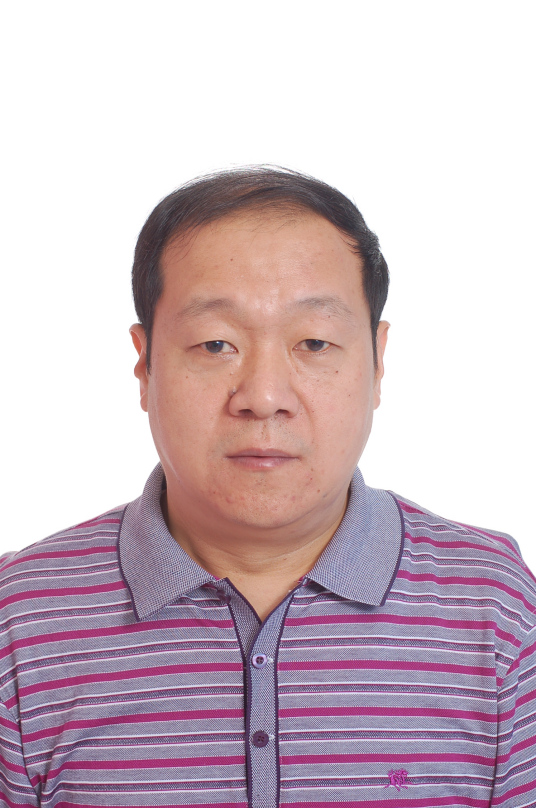}}]{Qing He}received the BS degree
        from Hebei Normal University, Shijiazhuang,
        China, in 1985, the MS degree from Zhengzhou
        University, Zhengzhou, China, in 1987, both in
        mathematics, and the PhD degree in fuzzy mathematics and artificial intelligence from Beijing Normal University, Beijing, China, in 2000. He is a
        professor as well as a doctoral tutor with the Institute of Computing Technology, Chinese Academy
        of Science (CAS), Beijing, China, and he is a professor with the University of Chinese Academy of
        Sciences (UCAS), Beijing, China. His interests include data mining,
        machine learning, classification, and fuzzy clustering.
        \end{IEEEbiography}

        \end{document}